\documentclass[final,5p,times,sort&compress]{elsarticle}

\usepackage{amssymb}
\usepackage{amsmath}
\usepackage{subcaption}
\usepackage[colorlinks=true,linkcolor=blue,citecolor=blue,urlcolor=blue]{hyperref}

\usepackage{lineno}

\journal{Nuclear Physics B}

\begin{document}

\begin{frontmatter}



\title{Measurement of spin correlation and entanglement in ATLAS and CMS}


\author{Fiona Ann Jolly$^{*}$, on behalf of the ATLAS and CMS Collaborations} 
\ead{fiona.ann.jolly@cern.ch}
\affiliation{organization={Deutsches Elektronen-Synchrotron DESY},
            addressline={Notkestrasse 85}, 
            postcode={22607}, 
            state={Hamburg},
            country={Germany}}

\begin{abstract}
The Large Hadron Collider has delivered exceptionally large proton--proton collision datasets at centre-of-mass energies of up to 13.6 TeV. These datasets, collected by the ATLAS and CMS detectors, enable precision tests of theoretical predictions using extensive samples of top-quark events. A key example is the study of top-quark pair spin correlations, which can be accessed through the angular distributions of the top-quark decay products owing to the unique property that the top quark decays before it can hadronise. This wealth of data has also enabled new measurements of top-quark pair properties, particularly those with enhanced sensitivity in the threshold region, such as quantum entanglement, that were previously out of reach. In this contribution, the latest highlights in this area from the ATLAS and CMS experiments, are presented.
\end{abstract}




\end{frontmatter}



\section{Introduction}
\label{sec:intro}
\begingroup
\renewcommand\thefootnote{*}
\footnotetext{Speaker, 18th International Workshop on Top Quark Physics (TOP2025)}
\endgroup
\begingroup
\renewcommand{\thefootnote}{}
\footnotetext{Copyright 2025 CERN for the benefit of the ATLAS and CMS Collaborations. CC-BY-4.0 license.}
\endgroup

With a mass of $172.69 \pm 0.30\,\text{GeV}$~\cite{10.1093/ptep/ptac097}, the top quark occupies a special place as the heaviest elementary particle in the Standard Model (SM). As a result, it has the strongest interaction with the Higgs boson, making it a key probe in theories of electroweak symmetry breaking and in studies of vacuum stability. Due to its extremely short lifetime ($\sim 10^{-25}$s)~\cite{PhysRevD.85.091104}, it decays before it can hadronise ($\sim 10^{-23}$s)~\cite{BIGI1986157}, transferring its spin information directly to its decay products. This unique feature provides a rare opportunity to probe the behaviour of an essentially unconfined quark.

At the Large Hadron Collider (LHC)~\cite{Lyndon_Evans_2008}, top quarks are predominantly produced in pairs ($t\bar{t}$) via gluon–gluon fusion. Their decay before hadronisation provides a unique opportunity to study phenomena such as spin correlations and quantum entanglement at TeV-scale energies. Unlike low-energy tests of quantum mechanics (QM) using photons, ions, or electrons~\cite{PhysRevLett.49.91,PhysRevLett.79.1,doi:10.1126/science.1130886,PhysicsPhysiqueFizika.1.195,PhysRev.47.777}, these high-energy proton--proton ($pp$) collisions allow the direct investigation of quantum effects in the quark sector with data from the ATLAS\footnote{\textbf{A} \textbf{T}oroidal \textbf{L}HC \textbf{A}pparatu\textbf{S}}~\cite{The_ATLAS_Collaboration_2008} and CMS\footnote{\textbf{C}ompact \textbf{M}uon \textbf{S}olenoid}~\cite{The_CMS_Collaboration_2008} experiments.

The most common $t\bar{t}$ decay channels are the dilepton channel, in which both top quarks decay semi-leptonically into a $b$-jet and a $W$ boson that subsequently decays to a charged lepton and a neutrino, and the lepton+jets channel, where one top quark decays hadronically into a $b$-jet and two light jets from the $W$-boson decay, while the other decays semi-leptonically. Measurements of $t\bar{t}$ spin correlations and entanglement in both the decay channels have been performed by the ATLAS~\cite{PhysRevLett.108.212001,PhysRevD.90.112016,PhysRevLett.114.142001,spincorr_2017,Aaboud_2020,Entanglement_2024} and CMS~\cite{PhysRevLett.112.182001,Spincorr_2016321,PhysRevD.93.052007,PhysRevD.100.072002,CMS_Collaboration_2024,PhysRevD.110.112016,CMS-PAS-TOP-25-001} collaborations. These measurements exploit the fact that the top-quark spin information is encoded in the angular distributions of its decay products. The following sections highlight recent results in this sector, assuming in all cases that the $t\bar{t}$ system is described by QM and the SM.

Since top quarks are spin-$\frac{1}{2}$ particles, each top quark can be treated as a two-level quantum system with two orthogonal spin states compared to a given axis, or a qubit in the language of quantum information theory. In QM, a quantum state is described by a density matrix, $\rho$, a non-negative operator in Hilbert space satisfying $\mathrm{tr}(\rho)=1$. For a bipartite Hilbert space given by the direct product of two subsystems, $a$ and $b$, a quantum state is called separable if:
\begin{equation}
\label{eq:separable_matrix}
    \rho = \sum_{n=1}^{N}p_{n}\rho_{n}^{a}\otimes\rho_{n}^{b}, \sum_{n=1}^{N}p_{n} = 1, p_{n}\geq 0,
\end{equation}
where $\rho_{n}^{a}$ and $\rho_{n}^{b}$ are quantum states of the sub-systems $a$ and $b$. The index $n$ labels the different product states in the mixture, each weighted with probability $p_{n}$. Any classically correlated state in the Hilbert space can be expressed in this form \cite{PhysRevA.40.4277}. 
The special case $N=1$ and $p_{n}=1$ corresponds to a pure product state. For $t\bar{t}$, however, one typically encounters mixed states.

Using this formalism, the spin-density matrix for a top-quark pair can be expressed in the standard basis of Pauli matrices as:
\begin{equation*}
    \label{spin_density_matrix}
        \rho=\frac{\mathbb{I}_4+\sum_{i=1}^{3}\left(B_i^{+}\sigma^{i}\otimes \mathbb{I}_2+B_i^{-}\mathbb{I}_2\otimes\sigma^{i}\right)+\sum_{i,j=1}^{3}C_{ij}\sigma^{i}\otimes\sigma^{j}}{4},
\end{equation*}
where $\mathbb{I}_n$ is the $n \times n$ identity matrix, $\sigma^i$ are the corresponding Pauli matrices, $B_i^{+} = \langle \sigma^{i}\otimes \mathbb{I}_2 \rangle$  and $B_i^{-} = \langle \mathbb{I}_2 \otimes \sigma^{i} \rangle$ denote the expectation values of the individual spin polarisations of the top and antitop quarks along the $i$-th axis in the chosen spin basis, and $C_{ij} = \langle \sigma^{i} \otimes \sigma^{j} \rangle$ encodes the spin correlations.
Measuring all 15 expectation values, $\langle \sigma^{i}\otimes \mathbb{I}_2 \rangle$, $\langle \mathbb{I}_2 \otimes \sigma^{i} \rangle$, and $\langle \sigma^{i} \otimes \sigma^{j} \rangle$, allows for the experimental reconstruction of the $t\bar{t}$ quantum state. In the SM, $C_{ij}$ is symmetric ($C_{ij}=C_{ji}$), which reduces the number of independent components, yet measuring all of them provides a stringent test of the SM.

Incorporating the spin-density matrix into the $t\bar{t}$ production cross-section, and considering the angular distributions of the decay products, the differential cross-section can be written as:
\begin{equation}
    \label{diff_xsec}
        \frac{1}{\sigma}\frac{d\sigma}{d\Omega_{1}\Omega_{2}}=\frac{1}{4\pi^{2}}\left( 1 + \alpha_{1}\textbf{B}_{1}\cdot\hat{e}_{1} + \alpha_{2}\textbf{B}_{2}\cdot\hat{e}_{2} + \alpha_{1}\alpha_{2}\hat{e}_{1}\cdot\textbf{C}\cdot\hat{e}_{2}\right),
\end{equation}
where $\hat{e}_{1}$ and $\hat{e}_{2}$ denote the directions of the top- and antitop-quark decay products in their respective rest frames, and $\Omega_{1,2}$ are the corresponding solid angles. 
The vectors $\mathrm{\textbf{B}}_{1,2}$ and the matrix $\mathbf{C}$ describe the top/antitop-quark spin polarisations and the spin correlation matrix, respectively. The quantities $\alpha_{1}$ and $\alpha_{2}$ are called the spin analysing powers of the decay products, which quantify how strongly the directions of the decay products reflect the parent top quark spin~\cite{BRANDENBURG2002235}. The spin analysing power ranges from -1 to 1, with $\alpha=\pm1$ indicating that the decay product is a perfect spin analyser and $\alpha=0$ indicating no spin sensitivity. For semi-leptonic top-quark decays, the lepton is the most sensitive spin analyser, with $|\alpha|=1$ at leading order (LO). For hadronic decays, the down-type quark jet coming from the $W$-boson decay carries maximal spin information ($|\alpha|=1$ at LO), though its identification is challenging and often relies on indirect methods, such as first identifying the up-type jet in a $W \rightarrow cs$ decay using algorithms that distinguish heavy-flavour decays within jets. The analysis in Section~\ref{sec:CMS_entanglement_boosted} employs strategies to reliably identify these jets.

Equation~\ref{diff_xsec} demonstrates that all 15 spin parameters of the top-quark pair can be extracted from the angular distributions of its decay products.

\section{Measurement of top quark polarisation and spin correlations using dilepton final states (CMS)}
\label{sec:CMS_spin}
CMS~\cite{PhysRevD.100.072002} measured all 15 spin-density matrix coefficients using dilepton events ($e^{+}e^{-}$, $e^{\pm}\mu^{\mp}$, $\mu^{+}\mu^{-}$) collected at centre-of-mass energy $\sqrt{s}=13\,\mathrm{TeV}$ in 2016, with $36\,\mathrm{fb}^{-1}$ of integrated luminosity.
    
The components are extracted using angular distributions defined in the helicity basis, which is an orthonormal basis constructed in the $t\bar{t}$ centre-of-mass frame. This basis consists of three quantisation axes, $\hat{k}$, $\hat{n}$, and $\hat{r}$, with respect to which spin correlation coefficients and polarisations are defined. Here, $\hat{k}$ is the top-quark direction of flight, $\hat{r}=( \hat{p}-\cos \Theta\hat{k})/\sin \Theta$ and $\hat{n}=\hat{r} \times \hat{k}$, where $\hat{p}$ is the beam direction and $\Theta$ is the top-quark production angle such that $\cos\Theta=\hat{k}\cdot\hat{p}$. The antitop-quark spin axes are defined analogously. 
    
\begin{figure}[htpb]
    \centering
    \includegraphics[width=0.35\textwidth]{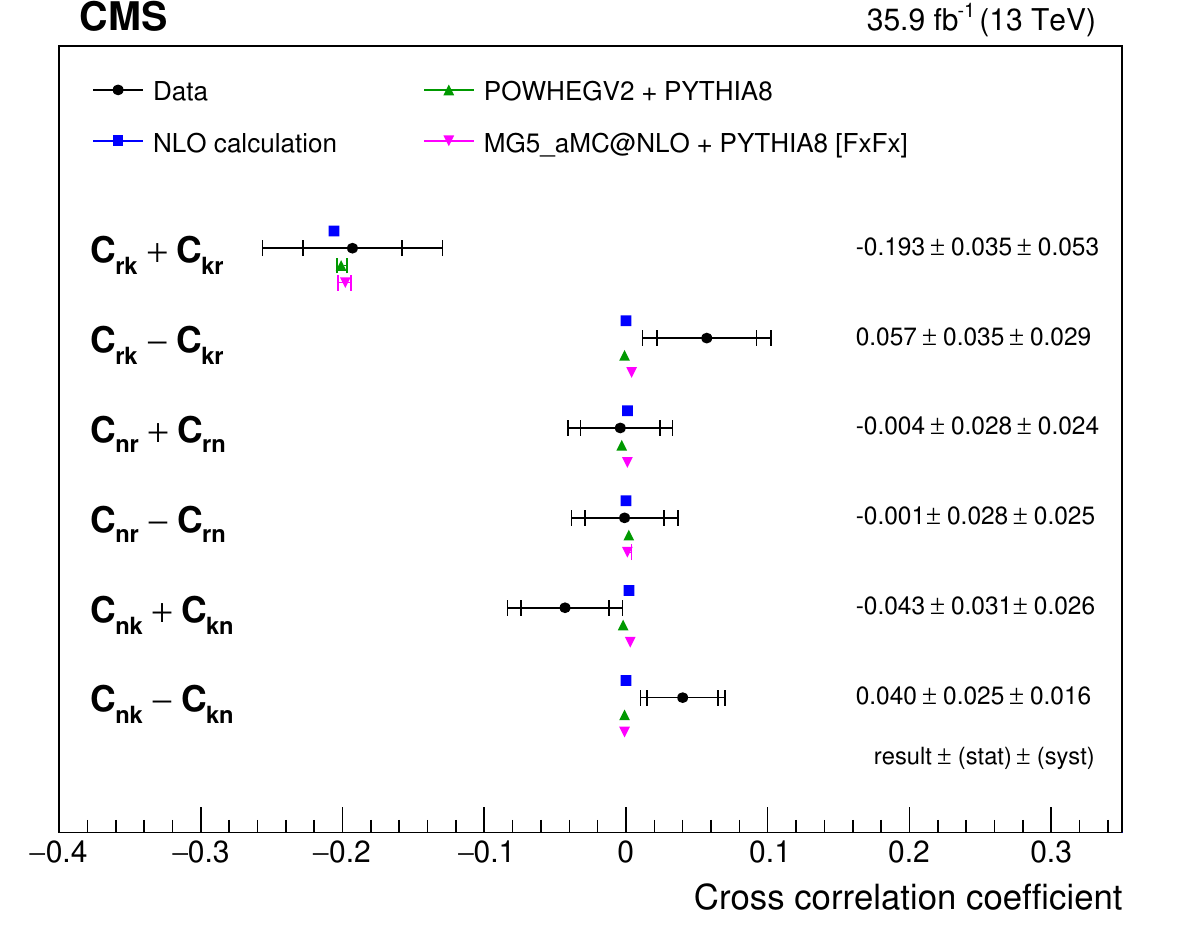}
    \caption{Measured spin correlation coefficients (circles) together with the nominal prediction using the Monte Carlo (MC) event generators POWHEG2 (triangles), an alternative sample from MADGRAPH5\_aMC@NLO (inverted triangles), and the fixed-order NLO calculation of Ref.~\cite{Bernreuther_2015} (squares). The inner vertical bars represent the statistical uncertainty and the outer bars the total uncertainty~\protect\cite{PhysRevD.100.072002}.}
    \label{fig:crosscorr}
\end{figure}

\begin{figure}[htpb]
    \centering
    \includegraphics[width=0.35\textwidth]{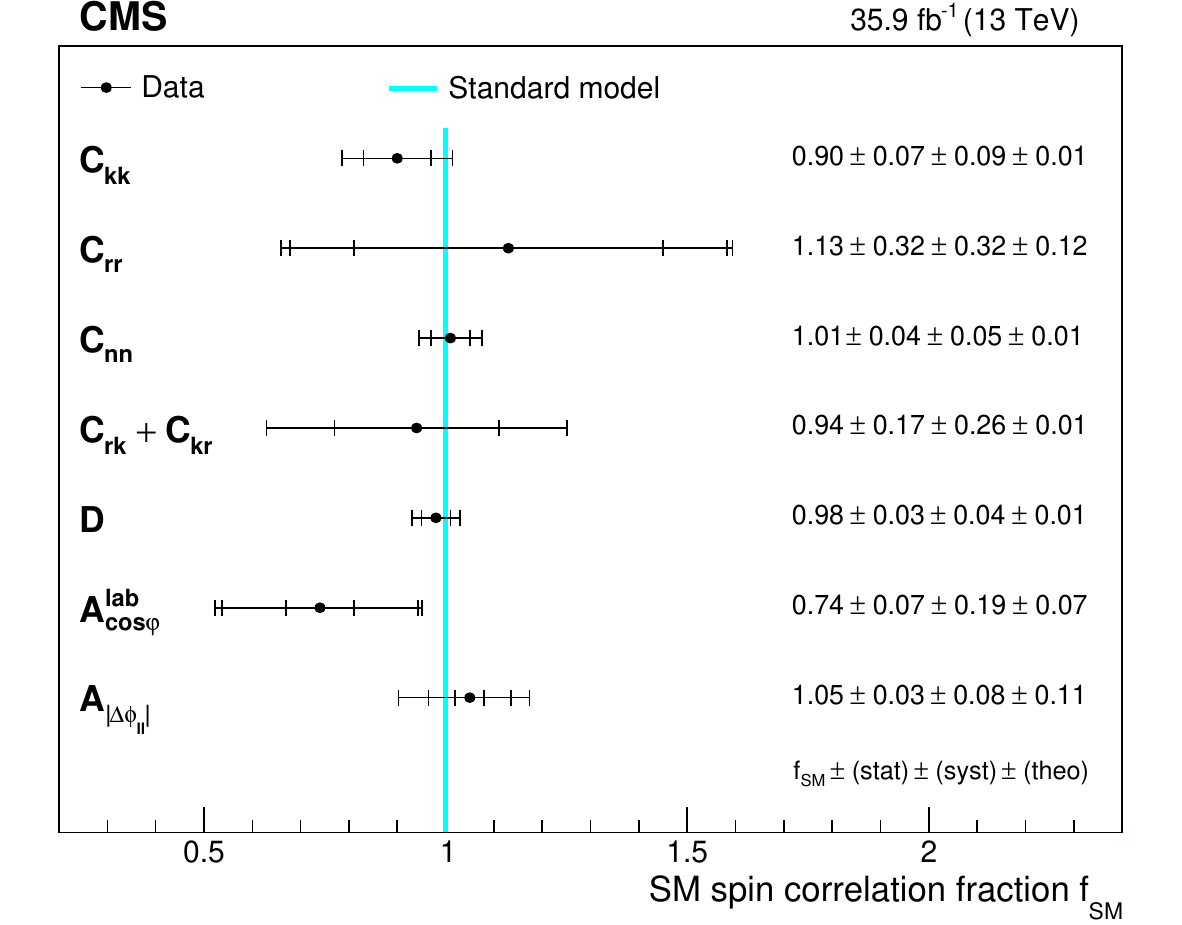}
    \caption{Measured values of \protect$f_{\mathrm{SM}}$\protect, the strength of the observed spin correlations relative to the SM prediction. The inner vertical bars indicate the statistical uncertainty, the middle bars the total experimental uncertainty, and the outer bars the total uncertainty~\protect\cite{PhysRevD.100.072002}.}
    \label{fig:fSM_CMS}
\end{figure}

A correlation coefficient $C_{ij}$ is determined from the distribution of the product $\cos\theta^{a}_{i}\cdot\cos\theta^{b}_{j}$, where $\cos\theta^{a}_{i}$ is the cosine of the angle between spin analyser $a$ (here, the charged lepton) and quantisation axis $i$. Each coefficient is thus extracted from a normalised parton-level differential cross-section that is specifically sensitive to it. Since detector resolution, acceptance, and reconstruction effects distort the observed angular distributions, the measured distributions are unfolded to parton level using the TUnfold regularised method~\cite{S_Schmitt_2012}. Figure~\ref{fig:crosscorr} shows a subset of the measured spin correlation coefficients. All coefficients agree with the SM expectations within uncertainties and are also consistent with the previous ATLAS full spin-density matrix measurement~\cite{spincorrelation_ATLAS_2017}. These coefficients are also used to extract the parameter $f_{\mathrm{SM}}$, defined as the fraction of SM-like spin correlation in the data. A value of $f_{\mathrm{SM}} = 1$ corresponds to the SM, while values above or below unity indicate a stronger or weaker degree of spin correlation relative to the SM. The extracted $f_{\mathrm{SM}}$ values, shown in Figure~\ref{fig:fSM_CMS}, are compatible with the SM. 

\section{ATLAS measurement of top-quark pair spin correlations}
\label{sec:ATLAS_spincorrelations}
The latest ATLAS measurement of spin correlations in top-quark pairs~\cite{Aaboud_2020} uses $36.1\,\mathrm{fb}^{-1}$ of $pp$ collision data collected in 2015--2016 at $\sqrt{s}=13\,\mathrm{TeV}$. The analysis extracts $f_{\mathrm{SM}}$ by fitting hypothesis templates to the parton-level unfolded normalised differential cross-sections. Unfolding is performed using an interative Bayesian method~\cite{DAGOSTINI1995487}. Two hypotheses are considered: dileptonic $t\bar{t}$ events with SM spin correlations, and dileptonic events with spin correlations removed. These templates are fitted to data via a binned maximum-likelihood fit implemented with MINUIT~\cite{Cacciari_2008}. The primary observable used in the fit is $\Delta\phi(l^{+},l^{-})$, the azimuthal separation between the charged leptons in the laboratory frame. This is the same observable used by CMS to derive the asymmetry $A_{|\Delta_{\phi}|}$ (see Figure~\ref{fig:fSM_CMS}).

Fits are also performed in bins of the $t\bar{t}$ invariant mass, $m_{t\bar{t}}$, with results summarised in Table~\ref{tab:fSM_ATLAS}. Overall, the measured spin correlation is slightly stronger than the SM prediction, with a deviation of about $2.2\sigma$.

\begin{table*}[htpb]
    \caption{Extracted $f_{\mathrm{SM}}$ values for each region with total uncertainties, along with the significance relative to the SM hypothesis. Significances account for statistical and systematic uncertainties in the data and theory uncertainties in the templates, including scale variations, parton density functions (PDFs), and the effect of top-quark decays. Values in parentheses show results excluding theory uncertainties~\cite{Aaboud_2020}.}
    \label{tab:fSM_ATLAS}
    \centering
    \begin{tabular}[htpb]{c|c|c}
      \hline
      Region & $f_{\mathrm{SM}}\pm(\mathrm{stat.,syst.,theory})$ & Significance (excl. theory) \\
      \hline
      Inclusive & $1.249\pm0.024\pm0.061^{+0.067}_{-0.090}$ & 2.2(3.8) \\
      \hline
      $m_{t\bar{t}}<450\,\mathrm{GeV}$  & $1.12\pm0.04^{+0.12}_{-0.13}{}^{+0.06}_{-0.07}$ & 0.78(0.87)\\
      $450\leq m_{t\bar{t}}<550\,\mathrm{GeV}$  & $1.18\pm0.08^{+0.13}_{-0.14}{}^{+0.13}_{-0.15}$ & 0.84(1.1)\\
      $550\leq m_{t\bar{t}}<800\,\mathrm{GeV}$  & $1.65\pm0.19^{+0.31}_{-0.41}{}^{+0.26}_{-0.33}$ & 1.2(1.4)\\
      $m_{t\bar{t}} \geq 800\,\mathrm{GeV}$  & $2.2\pm0.9^{+2.5}_{-1.7}{}^{+1.2}_{-1.5}$ & 0.49(0.61)\\
      \hline
     \end{tabular}
\end{table*}

\section{First observation of $t\bar{t}$ quantum entanglement (ATLAS)}
\label{sec:ATLAS_entanglement}
If a bipartite quantum state, as in Equation~\ref{eq:separable_matrix}, cannot be written as a separable product of its sub-systems, it is called entangled. Such states exhibit non-classical correlations, meaning the state of one particle cannot be fully determined independently of the other. The Peres-Horodecki criterion~\cite{HORODECKI1997333} provides a practical test for separability. It states that if a state is separable, the partial transpose with respect to one subsystem, $\rho^{\mathrm{T}_{b}}=\sum_{n=1}^{N}p_{n}\rho_{n}^{a}\otimes(\rho_{n}^{b})^{\mathrm{T}}$,  must be a valid physical state (i.e., a non-negative operator with unit trace). Negative eigenvalues of $\rho^{\mathrm{T}_{b}}$ indicate entanglement. Applying this criterion to the $t\bar{t}$ spin-density matrix yields a set of necessary conditions for entanglement. After averaging over all possible top-quark directions, entanglement survives only in two kinematic regimes: (i) close to production threshold, where the pair is predominantly a spin-singlet state, and (ii) at high energies with large scattering angles, where a spin-triplet dominates~\cite{Afik_2021}. Motivated by this, ATLAS performed the first measurement of quantum entanglement at the LHC~\cite{Entanglement_2024}, focusing on the threshold region, $340 <m_{t\bar{t}}<380\,\mathrm{GeV}$. Measurement in this narrow kinematic region is challenging due to limited number of events that satisfy the event selection criteria, the need for the accurate reconstruction of the $t\bar{t}$ system and identification of spin analysers, and theoretical modelling uncertainties. The analysis used dilepton final states from $pp$ collisions at $\sqrt{s}=13\,\mathrm{TeV}$ with $140\,\mathrm{fb}^{-1}$ of integrated luminosity. In this region, if the trace of the spin correlation matrix 
satisfies $\mathrm{tr}[\mathbf{C}]+1 < 0$, entanglement is indicated. This condition can be expressed in terms of an entanglement witness, $D=\mathrm{tr}[\mathbf{C}]/3$, which can be experimentally extracted via $ D = 3 \cdot \frac{\langle \cos\varphi \rangle}{\alpha_{1}\alpha_{2}}$. Here, $\varphi$ is the angle between the spin analysers in the top- and antitop-quark rest frames, and $|\alpha_{1,2}|=1$. The entanglement criterion then becomes  $D< -\frac{1}{3}$. To determine $D$, ATLAS employed a calibration-curve method that maps the reconstructed value of $D$ to its particle-level value, defined using stable particles after hadronisation but before detector effects. The curve is constructed by reweighting simulated samples at both detector and particle levels, using parton-level information to generate a range $D$ values. 

\begin{figure}[htpb]
    \centering
        \includegraphics[width=0.4\textwidth]{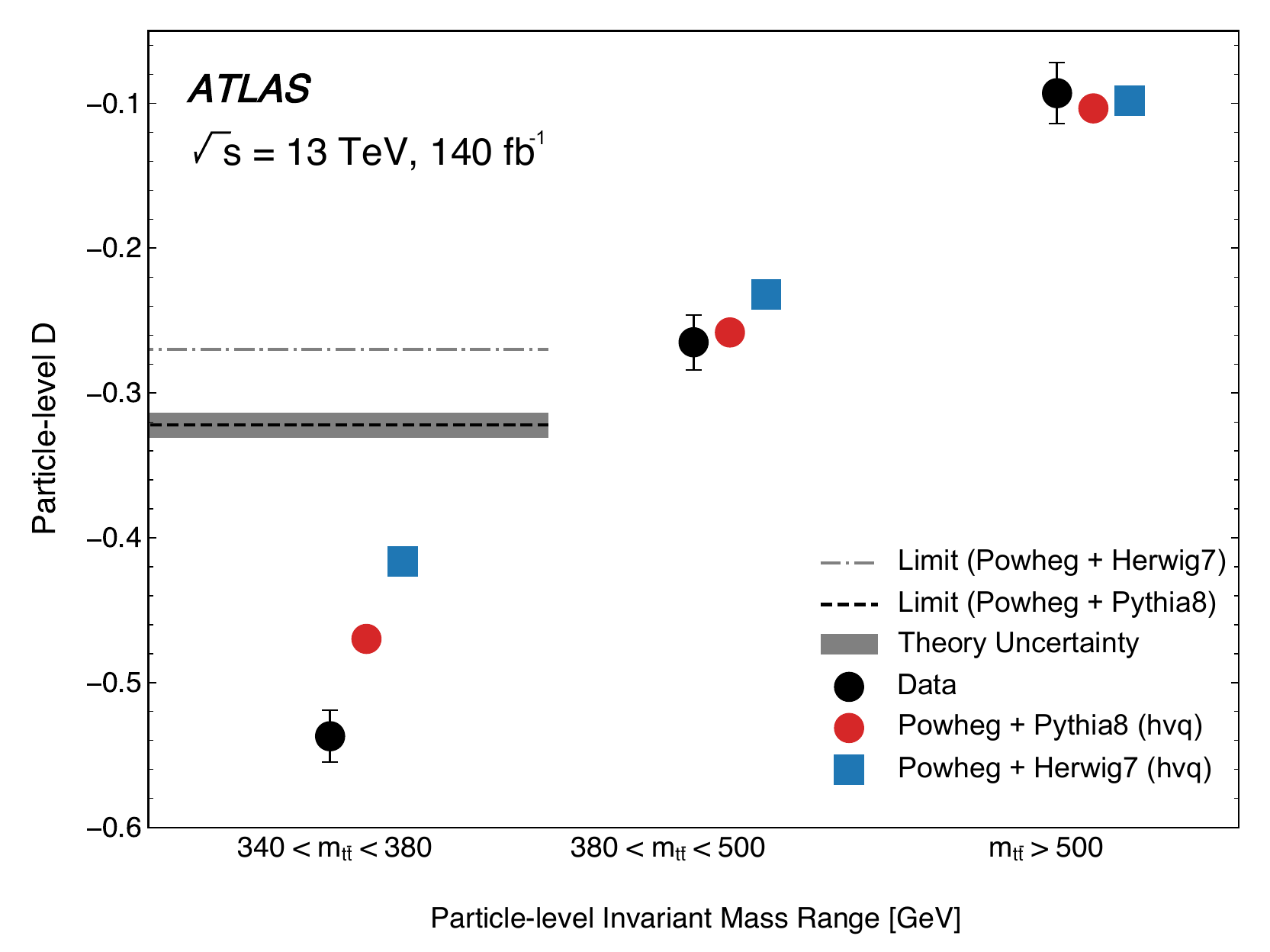}
    \caption{The particle-level $D$ results in the signal and validation regions are compared with various MC models. The entanglement limit, converted from the parton-level value $D=-1/3$ to particle level, is shown~\cite{Entanglement_2024}.}%
    \label{fig:ATLAS_entanglement}
\end{figure}

Figure~\ref{fig:ATLAS_entanglement} shows the measured value, $D=-0.537 \pm 0.002 (\mathrm{stat.}) \pm 0.019 (\mathrm{syst.})$, together with the SM predictions for the threshold and two validation regions. The entanglement limit, $D=-1/3$, is propagated using a similar curve from parton to particle level. In the threshold region, the measured $D$ lies well below the separability bound, indicating entanglement with a significance exceeding $5\sigma$. The dominant systematic uncertainty arises from parton-shower modelling, evaluated by comparing the nominal \textsc{Powheg+Pythia} sample with \textsc{Powheg+Herwig}. Notably, the difference between these generators is the largest in the threshold region. Furthermore, a tension between the data and SM predictions is observed in this region, a discrepancy that is not present at higher masses. This suggests that current modelling does not fully capture the dynamics of $t\bar{t}$ production near threshold. While this ATLAS result provides the first clear evidence of entanglement in $t\bar{t}$, a more complete theoretical description of the threshold regime is required. 
In particular, effects from $t\bar{t}$ quasi-bound states, accounted for in the CMS entanglement analysis via a simplified model (Section~\ref{sec:CMS_entanglement}), are expected to play an important role.

\section{Observation of quantum entanglement by CMS}
\label{sec:CMS_entanglement}
\begin{figure}[htpb]
    \centering
        \includegraphics[width=0.4\textwidth]{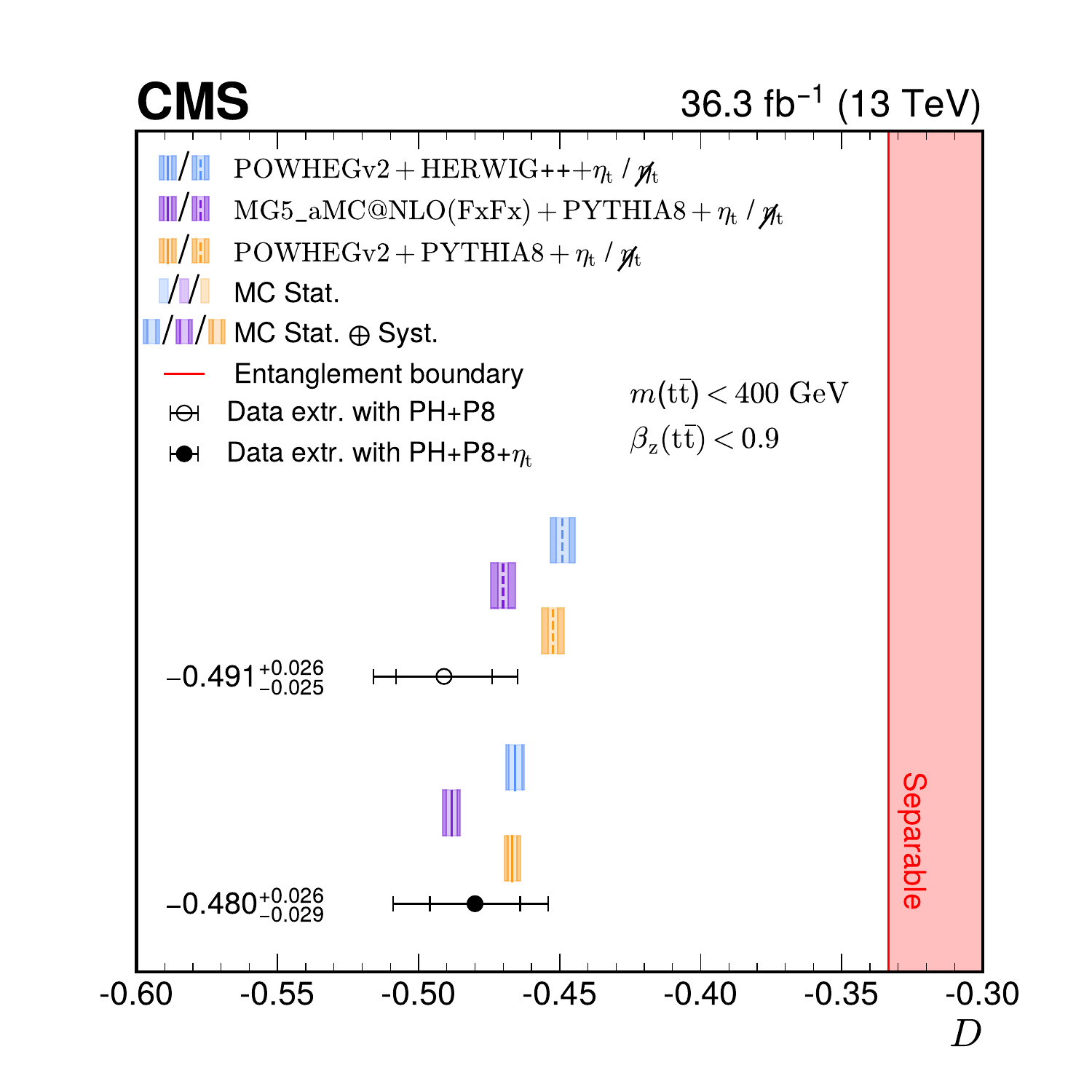}
    \caption{Summary of the measured entanglement proxy $D$ (black points) compared with MC predictions with (solid) and without (dashed) the $\eta_t$ state. Inner (outer) error bars show the statistical (total) uncertainty in data. Light (dark) shaded regions indicate the MC statistical (total) uncertainty. The entanglement limit $D=-1/3$ is shown as a shaded band~\cite{CMS_Collaboration_2024}.}%
    \label{fig:CMS_entanglement}
\end{figure}

Based on $pp$ collisions data from 2016 at $\sqrt{s}=13 \, \mathrm{TeV}$ with an integrated luminosity of $36.3 \, \mathrm{fb}^{-1}$, CMS measured entanglement in the $t\bar{t}$ dilepton channel near threshold~\cite{CMS_Collaboration_2024}. The region is defined by $345 < m_{t\bar{t}} < 400 \, \mathrm{GeV}$ and $\beta_z({t\bar{t}}) < 0.9$, where $\beta_z({t\bar{t}})$ is the relative longitudinal velocity between the laboratory and $t\bar{t}$ reference frames. This selection on $\beta_z({t\bar{t}})$ enhances the fraction of gluon–gluon fusion events~\cite{Aguilar_Saavedra_2022}, in which the singlet configuration dominates. The observable $D$ is extracted at parton level via a binned profile likelihood fit to the $\cos \varphi$ distribution, incorporating, for the first time in an LHC measurement, simplified $t\bar{t}$ quasi-bound state effects. Such quasi-bound effects were first predicted in the late 1980s~\cite{Fadin:1987wz,Fadin:1990wx,PhysRevD.47.56}, nearly a decade before the discovery of the top quark at the Tevatron in 1995~\cite{D0:1995jca,CDF:1995wbb}. In this measurement, these effects are modelled as a pseudoscalar resonance $\eta_{t}$~\cite{maltoni2024quantumdetectionnewphysics}. Figure~\ref{fig:CMS_entanglement} shows the fitted and predicted values of $D$ with and without the $\eta_{t}$ contribution. While entanglement is observed with a significance exceeding $5\sigma$ in both with and without $\eta_{t}$ scenarios, the data is better described when $\eta_{t}$ is included in the signal model. The largest impact on the fit arises from the 50\% normalisation applied to $\eta_{t}$ in the fit. Despite the simplified modelling, this analysis represents a step towards accounting for dynamics at threshold. Ongoing theoretical studies aim to refine these descriptions~\cite{Sumino_2010,Hagiwara_2008,Fuks_2025}, and the threshold region has become increasingly important in light of recent ATLAS~\cite{ATLAS-CONF-2025-008} and CMS~\cite{Collaboration_2025} results where an excess compatible with a quasi–bound-state formation at threshold has been observed in both experiments.

\section{Measurements of polarisation, spin correlations, and entanglement using lepton+jets events}
\label{sec:CMS_entanglement_boosted}
CMS has extended measurements of top-quark spin phenomena to the lepton+jets channel, using the full Run~2 dataset of $pp$ collisions at $\sqrt{s}=13 \,\mathrm{TeV}$ corresponding to $138 \,\mathrm{fb}^{-1}$. All coefficients of the spin-density matrix are extracted simultaneously through a binned maximum-likelihood fit~\cite{PhysRevD.110.112016}. The analysis is performed inclusively and differentially as a function of $m_{t\bar{t}}$, $|\cos(\theta)|$, and the top-quark transverse momentum $p_{\text{T}}(t)$, where $\theta$ is the production angle.

\begin{figure*}[htpb]
    \centering
        \includegraphics[width=0.7\textwidth]{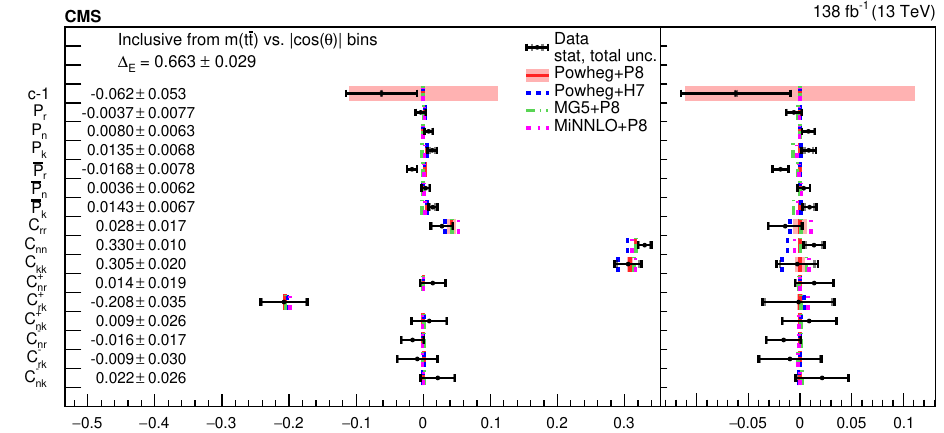}
    \caption{Results of the inclusive full matrix measurement. The measurements (markers) are shown with statistical (inner) and total (outer) uncertainties and compared to predictions from POWHEG+PYTHIA, POWHEG+Herwig, MadGraph5\_aMC@NLO+PYTHIA, and MiNNLO+PYTHIA. Right panels show results with the POWHEG+PYTHIA prediction subtracted. The corresponding $\Delta_E$ values are indicated~\cite{PhysRevD.110.112016}.}%
    \label{fig:spin_ljets}
\end{figure*}

For the hadronic top-quark decay, the down-type quark jet serves as the spin analyser, with a spin analysing power of unity at LO. Correctly identifying this jet among the final-state jets is challenging. To address this, an artificial neural network is employed that exploits information from algorithms designed to distinguish heavy-flavour decays within jets, which enhances the probability of correctly identifying a $c$-jet from the $W$-boson decay and assigning the remaining jet as the down-type jet. The extracted inclusive spin-density matrix coefficients, shown in Figure~\ref{fig:spin_ljets}, are in good agreement with SM predictions.

\begin{figure}[htpb]
    \centering
        \includegraphics[width=0.4\textwidth]{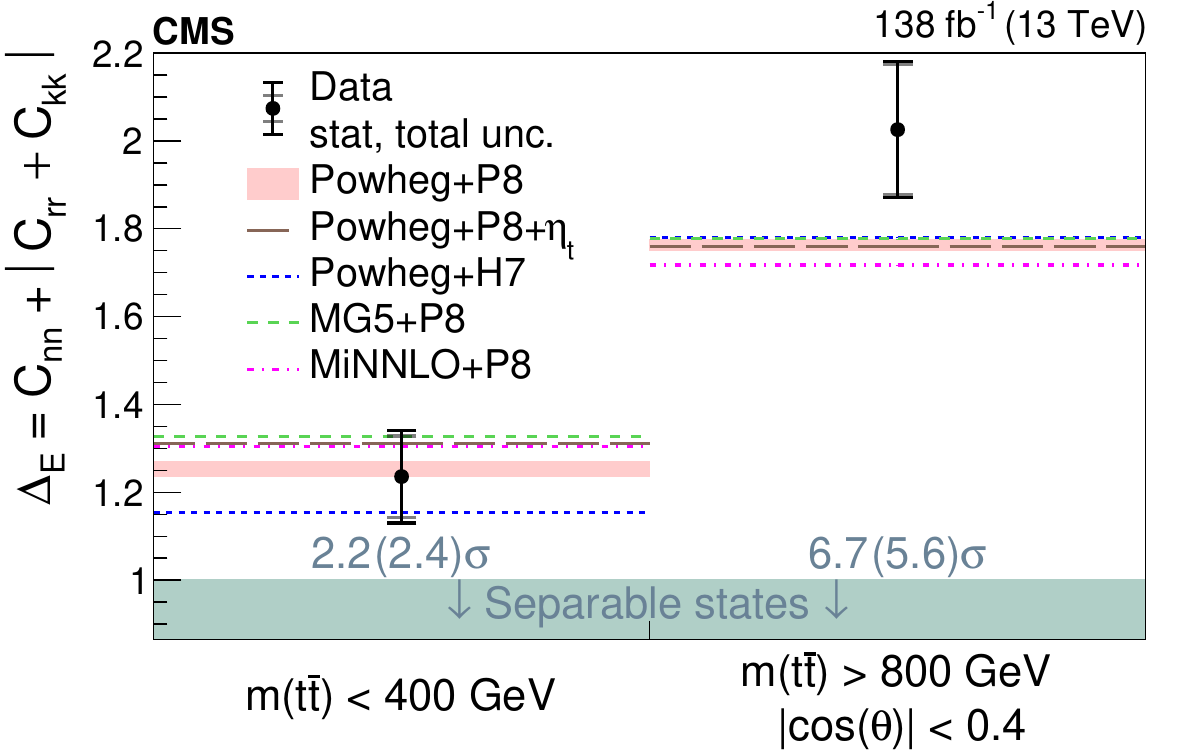}
    \caption{Entanglement measurement results using the full matrix in various $m_{t\bar{t}}$ regions. Points show statistical (inner) and total (outer) uncertainties, compared to POWHEG+PYTHIA, POWHEG+PYTHIA+$\eta_t$, POWHEG+Herwig, MadGraph5\_aMC@NLO+PYTHIA, and MiNNLO+PYTHIA predictions. POWHEG+PYTHIA includes matrix element (from the hard interaction of partons) scale and PDF uncertainties; others show central values only. Observed (expected) significance relative to separable states (green band) is given in $\sigma$~\cite{PhysRevD.110.112016}.}%
    \label{fig:entanglement_ljets}
\end{figure}

CMS also performed measurements of quantum entanglement, including for the first time the high-$m_{t\bar{t}}$ (boosted) regime. Using the Peres–Horodecki criterion, an observable analogous to the threshold witness is defined as $ \tilde{D} = 3 \cdot \frac{\langle \cos\chi \rangle}{\alpha_{1}\alpha_{2}}$, where $\chi$ differs from $\varphi$ by a sign flip of the $\hat{n}$-axis component of one of the spin analysers~\cite{Aguilar_Saavedra_2022}. Entanglement is indicated by $\tilde{D}>1/3$. A complementary approach uses full $t\bar{t}$ spin-density matrix reconstruction to define an entanglement-sensitive quantity $\Delta_{E} = C_{nn}+|C_{rr}+C_{kk}|$, with $\Delta_E>1$ signalling entanglement. In the boosted phase space defined by $m_{t\bar{t}}>800 \,\mathrm{GeV}$ and $|\cos(\theta)|<0.4$, the matrix-based observable provides greater sensitivity than $\tilde{D}$, as illustrated in Figure~\ref{fig:entanglement_ljets}. Owing to the limited event yield in this challenging region, the statistical uncertainty remains comparable in size to the systematic uncertainty. Nevertheless, this measurement clearly demonstrates that entanglement can be probed in the boosted regime, marking an important step towards studying high-energy $t\bar{t}$ spin correlations.

\section{Conclusion}
\label{sec:Conclusion}
Recent measurements of spin correlation and quantum entanglement in top-quark pairs at the highest LHC energies by the ATLAS and CMS collaborations have been presented. Spin polarisations and correlations are also related to other quantum properties, such as Magic~\cite{aoude2025probingnewphysicssector,white2024magicentangledquarks}, which quantifies the potential computational advantage of quantum states over classical states. CMS has measured this quantity in top-quark pairs and found agreement with the SM~\cite{CMS-PAS-TOP-25-001}. Entanglement is observed with a significance of more than $5\sigma$ at the $t\bar{t}$ threshold, although discrepancies between data and SM predictions indicate the need for further studies in this challenging region of phase space. CMS has additionally performed the first measurement of entanglement in the boosted regime, enabling exploration of quantum effects that may dominate at high $m_{t\bar{t}}$. One such effect is potential violation of Bell inequality~\cite{han2023quantumentanglementbellinequality}. While a full Bell test is not feasible in collider environments, since spin correlations can only be inferred from angular distributions and not measured event-by-event, studying observables sensitive to such violations provides valuable insight into the high-energy dynamics of top-quark pairs. These results mark a significant step towards probing quantum properties in top-quark systems and pave the way for comprehensive quantum information studies of top quarks at the LHC.



\bibliographystyle{elsarticle-num}
\bibliography{doc/proceedings}






\end{document}